\documentclass[prl,twocolumn,superscriptaddress,showpacs,amscd,amsmath,amssymb,verbatim]{revtex4}

\usepackage{graphicx}
\usepackage[dvips]{hyperref}
\usepackage{textcomp}
\usepackage[OT1,T1]{fontenc}

\begin{document}
\newcommand{\FeLang}{BNFSO}

\title{Role of Antisymmetric Exchange in Selecting Magnetic Chirality in Ba$_3$NbFe$_3$Si$_2$O$_{14}$}
\author{A. Zorko}
\affiliation{Jo\v{z}ef Stefan Institute, Jamova 39, SI-1000 Ljubljana, Slovenia}
\affiliation{EN--FIST Centre of Excellence, Dunajska 156, SI-1000 Ljubljana, Slovenia}
\author{M. Pregelj}
\affiliation{Jo\v{z}ef Stefan Institute, Jamova 39, SI-1000 Ljubljana, Slovenia}
\author{A. Poto\v{c}nik}
\affiliation{Jo\v{z}ef Stefan Institute, Jamova 39, SI-1000 Ljubljana, Slovenia}
\author{J. van Tol}
\affiliation{National High Magnetic Field Laboratory, Florida State University, Tallahassee, Florida 32310}
\author{A. Ozarowski}
\affiliation{National High Magnetic Field Laboratory, Florida State University, Tallahassee, Florida 32310}
\author{V. Simonet}
\affiliation{Institut N\'eel, CNRS and Universit\'e Joseph Fourier, BP 166, 38042 Grenoble, France}
\author{P. Lejay}
\affiliation{Institut N\'eel, CNRS and Universit\'e Joseph Fourier, BP 166, 38042 Grenoble, France}
\author{S. Petit}
\affiliation{Laboratoire L\'eon Brillouin, CEA-CNRS, CE-Sacley, F-91191 Gif sur Yvette, France}
\author{R. Ballou}
\affiliation{Institut N\'eel, CNRS and Universit\'e Joseph Fourier, BP 166, 38042 Grenoble, France}

\date{\today}
\begin{abstract}
We present electron spin resonance (ESR) investigation of the acentric Ba$_3$NbFe$_3$Si$_2$O$_{14}$, featuring a unique single-domain double-chiral magnetic ground state. Combining simulations of the ESR line-width anisotropy and the antiferromagnetic-resonance modes allows us to single out the Dzyaloshinsky-Moriya (DM) interaction as the leading magnetic anisotropy term. We demonstrate that the rather minute out-of-plane DM component $d_c=45$~mK is responsible for selecting a unique ground state, which endures thermal fluctuations up to astonishingly high temperatures. 
\end{abstract}
\maketitle

Spin-orbit coupling generally has no relevant effect in ordinary magnets, except to pin the orientation of spins with respect to the background lattice in magnetically ordered phases, singled out by the isotropic Heisenberg exchange interactions. An exception is provided on geometrically frustrated lattices where the isotropic interactions alone are often unable to raise the macroscopic degeneracy of a ground state (GS) \cite{LMM}, leading to unconventional cooperative electronic states \cite{LMM,Balents,Bramwell,ZorkoCuNCN}. Then, the magnetic behavior might be exclusively driven by minute perturbing terms in the form of anisotropic interactions emanating from the spin-orbit coupling. For instance, the antisymmetric Dzyaloshinsky-Moriya (DM) exchange interaction \cite{Moriya} can induce order on a kagom\'e lattice in the classical limit \cite{Elhajal,Ballou,Matan} and can lead to quantum criticality in the quantum limit \cite{Cepas}. Furthermore, DM is known to alter phase diagrams of frustrated ladders \cite{Penc} and triangular lattices \cite{Starykh}. It is also responsible for spin chirality \cite{MnSi} that can be long-ranged even in the absence of a classical magnetic order \cite{Grohol}. Spin chirality is one of the key concepts in the physics of strongly correlated electrons, as it is related to various intriguing phenomena, like the realization of spin liquids \cite{Machida}, magnetic-order induced ferroelectricity \cite{Mostovoy}, anomalous Hall effect \cite{Taguchi} and  possibly high-temperature superconductivity \cite{Wen}.

\begin{figure}[b]
\includegraphics[width=1\linewidth, trim=0 10 0 10, clip=true]{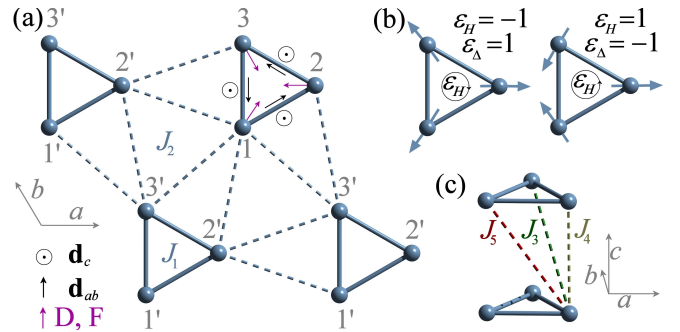}
\caption{(color online). 2D "triangular" arrangement of Fe$^{3+}�$ in \FeLang. (a),~(c) Exchange interactions $J_{1-5}$ and magnetic anisotropies of the Dzyaloshinsky-Moriya ({\bf d}) and single-ion ({\bf D}, {\bf F}) type. The latter vectors are parallel to a local two-fold rotational axis defining the $Z$ axis. (b) Two possible double chiral ground states, with $\epsilon_{\Delta}$ and $\epsilon_H$ denoting triangular chirality and helicity (with $\sim 2\pi/7$ pitch along $c$ axis), respectively.}
\label{fig1}
\end{figure}

In this context, Ba$_3$NbFe$_3$Si$_2$O$_{14}$~(BNFSO) is extremely appealing due to its remarkable magnetic properties. It crystallizes in a non-centrosymmetric trigonal unit cell (P321 symmetry). The Fe$^{3+}$ ($S=5/2$) spins reside on vertices of equilateral triangles arranged into a two dimensional (2D) triangular lattice (crystallographic $ab$ planes in Fig.~\ref{fig1}) \cite{MartyPRL}. The dominant exchange interactions are antiferromagnetic -- the Curie-Weiss temperature is $\theta\sim -180$~K \cite{MartyPRL,MartyPRB,ZhouCM} -- and thus frustrated. Nevertheless, a long-range-ordered (LRO) state is realized below $T_N=26$~K, characterized by a $120^{\circ}$ spin arrangement on each triangle. The moments are bound to the $ab$ planes and form a  magnetic helix along the crystallographic $c$ axis, corresponding to the magnetic propagation vector ${\bf q}=(0,0,\tau)$, $\tau \sim 1/7$ \cite{MartyPRL}. 

The system has been drawing considerable attention because of its magnetoelectric and multiferroic properties \cite{MartyPRB,ZhouCM,Lee}. However, its uniqueness stems from its distinctive magnetic order and magnetic excitations. The LRO magnetic GS is doubly chiral [Fig.~\ref{fig1}(b)] and single-domain -- a single triangular vector chirality $\epsilon_{\Delta}$ together with a single helicity $\epsilon_{H}$ is chosen from four possible states in a structurally chiral crystal -- which is believed to be crucial for its multiferroic properties \cite{MartyPRL}. Moreover, below $T_N$ one of the two magnetic excitation branches emerging from the magnetic satellites is completely chiral over the whole energy spectrum, implying an unprecedented absence of chirality mixing of the spin dynamics \cite{Loire}. The chiral correlations remain present far above $T_N$ \cite{Stock}. The helicity of the GS can be rationalized within the isotropic Heisenberg model if three competing interplane interactions [$J_{3-5}$ in Fig.~\ref{fig1}(c)] are considered. However, the isotropic exchange still allows \cite{MartyPRL} for two $(\epsilon_H,\epsilon_{\Delta})$ pairs [Fig.~\ref{fig1}(b)], and fails to justify the selection of the $ab$ easy plane. Thus, the fundamental question about the mechanism, responsible for the experimentally observed unique single-domain chiral GS \cite{MartyPRL} in BNFSO, arises. 

It was suggested that the DM anisotropy might be responsible for selecting the chiral GS \cite{MartyPRL} and for opening a small gap in magnetic excitations \cite{Loire} of BNFSO. In another inelastic neutron scattering (INS) study the latter was proposed to arise from single-ion (SI) anisotropy \cite{Stock}. A clear and quantitative determination of the source of the magnetic anisotropy and of its impact on the GS is therefore needed. In this Letter, we present our electron spin resonance (ESR) study, which provides a direct insight to magnetic anisotropy of BNFSO. Jointly simulating the experimental ESR line-width anisotropy in the paramagnetic phase and magnetic excitations in the ordered phase, we determine the dominant magnetic anisotropy, which is of the DM type and has sizable both the out-of-plane ($d_c$) and the in-plane ($d_{ab}$) component (Fig.~\ref{fig1}). Our mean-field based calculations single out $d_c$ as being responsible for selecting the unique GS. 

ESR is an extremely powerful technique for quantifying magnetic anisotropy \cite{AB}. The isotropic Heisenberg exchange commutes with the $S_z$ spin operator ($z$ denotes the quantization axis set by the applied magnetic field) and therefore leads to a $\delta$-function resonance. Finite magnetic anisotropy then yields a finite ESR line width. In single crystals the line-width anisotropy in the paramagnetic phase can often unveil the dominant magnetic anisotropy, since different types of anisotropy reflect local symmetries distinctively, as they arise from different microscopic origins. Finally, ESR can also detect collective magnon modes in a LRO state. We therefore performed an extensive ESR investigation on high-quality BNFSO single crystals in the temperature range between 4~K and 500~K. X-band (at 9.3 GHz) spectra were recorded on a home-build double-cavity spectrometer, equipped with a helium-flow cooling and a preheated-nitrogen-flow heating system. Measurements at frequencies between 50 GHz and 400 GHz were performed on a couple of custom-made transmission-type spectrometers at NHMFL, Florida.

\begin{figure}[t]
\includegraphics[width=1\linewidth, trim=0 0 0 22, clip=true]{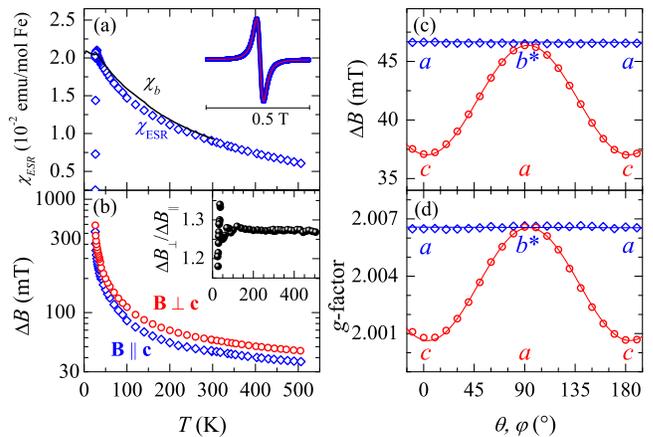}
\caption{(color online). X-band ESR results: (a) Comparison of ESR, $\chi_{\rm ESR}$, and bulk, $\chi_b$, susceptibility for $\bf{B}\|\bf{c}$. Inset: room-temperature ESR spectrum (blue) and the corresponding Lorentzian fit (red). (b) Temperature dependence of line widths and the line width ratio (inset). Angular dependence of (c) the line width and (d) the $g$-factor measured at 500~K, with corresponding fits (solid lines) explained in the text.}
\label{fig2}
\end{figure}

In X-band, Lorentzian-shaped spectra are observed down to $T_N$ (Fig.~\ref{fig2}), in accord with strong exchange narrowing regularly encountered in dense magnetic systems. Below $T_N$, the paramagnetic signal rapidly disappears. Our calibration of the ESR intensity, which is proportional to spin-only magnetic susceptibility $\chi_{\rm ESR}$, and the scaling between $\chi_{\rm ESR}$ and the bulk susceptibility $\chi_{b}$ [Fig.~\ref{fig2}(b)] prove that the ESR signal of BNFSO is intrinsic. The observed small deviations of the $g$-factor from the free-electron value $g_0=2.0023$ are typical for Fe$^{3+}$ with a nearly pure $^6$S$_{5/2}$ orbital singlet GS \cite{AB}. A uniaxial anisotropy of the $g$-tensor and the line width [Fig.~\ref{fig2}(c-d)] are justified by a three-fold rotational symmetry of iron triangles. Our fit of the $g$-factor anisotropy \cite{sup} yields the three eigenvalues of the $g$-tensor: $g_{XX}=2.000$, $g_{YY}=2.001$ and $g_{ZZ}=2.013$, and thus reveals that also locally the symmetry of the $g$-tensor is very close to being uniaxial. The polar axis $Z$ on each site is set by a local two-fold rotational axis lying within the crystallographic $ab$ plane ($a$ direction for site 2 in Fig.~\ref{fig1}). The other two principal axes $X$ and $Y$ for site 2 lie $30^{\circ}$ and 120$^{\circ}$ from the $b^*$ axis in the crystallographic $b^*c$ plane \cite{sup}.

The anisotropy of the $g$-factor reflects mixing of excited orbital states into the ground orbital singlet of the Fe$^{3+}$ ion, which is induced by a spin-orbit coupling. The anisotropy of the line width arises from magnetic anisotropy present in a spin Hamiltonian, which originates from the same perturbation. We address the issue of the magnetic anisotropy in BNFSO in the framework of the spin Hamiltonian
\begin{equation}
\label{eq1}
\mathcal{H}=  \sum _{(ij)}J_{ij} {{\bf S}_{i} \cdot {\bf S}_{j}} - \mu_{B}{\bf B}\cdot {\bf g} \cdot \sum_{j} {\bf S}_j + \mathcal{H'},
\end{equation}
\noindent where, the first sum runs oven the spin pairs connected by one of the five different exchange interactions (Fig.~\ref{fig1}) and represents the Heisenberg term $\mathcal{H}_e$, the second sum gives the Zeeman coupling $\mathcal{H}_Z$ and the third sum the magnetic anisotropy $\mathcal{H}'$. The two dominant contributions \cite{foot1} to the latter for Fe$^{3+}$ are the single-ion anisotropy \cite{AB}
\begin{eqnarray}
\label{eq2}
\mathcal{H'}_{\rm SI}&=& \sum_{j}  \bigg[ D S_{j,Z}^2 + \frac{a}{6} \left(S_{j,X}^4+S_{j,Y}^4+S_{j,Z}^4\right) + \nonumber\\
& + & \frac{F}{180} \left(35 S_{j,Z}^4 -\frac{475}{2} S_{j,Z}^2\right) \bigg] + const.,
\end{eqnarray}
\noindent and the Dzyaloshinsky-Moriya anisotropy \cite{Moriya}
\begin{equation}
\label{eq3}
\mathcal{H'}_{\rm DM} = \sum _{(ij)}{\bf d}_{ij}\cdot {{\bf S}_{i} \times {\bf S}_{j}}.
\end{equation}
\noindent The symmetry of the SI anisotropy is the same as of the $g$-tensor \cite{Gatteshi}, therefore we omit the additional $E\left(S_X^2-S_Y^2\right)$ term. Since $J_1$ is dominant \cite{Loire,Stock}, we consider the DM interaction only between nearest neighbors. The DM vector may possess a non-zero out-of-plane $d_c$ and in-plane $d_{ab}$ component parallel to the bond [Fig.\ref{fig1}(a)]. The third component is forbidden by the two-fold rotational axis passing through the middle of each bond. 

Using the Kubo-Tomita approach \cite{KT}, we can analytically calculate the line-width anisotropy $\Delta B_{\infty}(\theta)$ at infinite temperature, separately for both types of magnetic anisotropy \cite{sup}. At finite temperatures $\Delta B(\theta)=c(\theta,T)\Delta B_{\infty}(\theta)$, where $c(\theta,T)$ differs from unity due to finite spin correlations. In BNFSO the uncorrelated paramagnetic state is not reached yet even at 500~K ($\sim 3\theta$) [Fig.~\ref{fig2}(b)], which complements  specific heat measurements showing that $20\%$ of magnetic entropy is still missing at 200~K \cite{ZhouCM}. This needs to be contrasted with several Cu-based 2D frustrated lattices, where the ESR line width was found constant for $T\gtrsim \theta$ \cite{ZorkoSCBO,ZorkoHerb}.

\begin{figure}[t]
\includegraphics[width=1\linewidth, trim=0 6 0 16, clip=true]{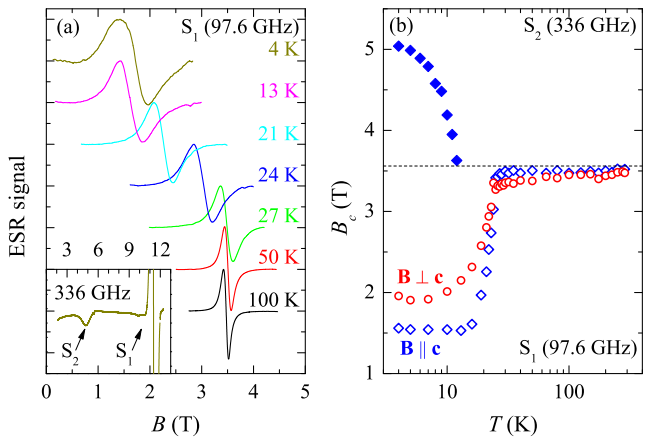}
\caption{(color online). (a) Selected ESR spectra at 97.6~GHz. Inset shows two distinct resonance modes at 4 K. (b) Temperature dependence of the central field of the two observed modes.}
\label{fig3}
\end{figure}

We find the fits of the experimental line width [Fig.~\ref{fig2}(c)] virtually indistinguishable for the SI and the DM  model. At 500~K we extract $\sqrt{c} |D| =0.53$~K, $a/D =1.06$, $F/D =0.09$ for the SI model and $\sqrt{c} |d_c| =0.13$~K, $d_{ab}/d_c =2.6$ for the DM model. It is worth noting that since Fe$^{3+}$ is in the orbital $S$ state, in the SI model only the $a$ term is allowed in purely cubic, octahedral or tetrahedral environments. In BNFSO the local symmetry is lower, therefore, $D$ and $a$ can have similar magnitudes  \cite{AB}. Although the absolute size or the sign of the anisotropy terms cannot be revealed from these fits due to unknown $c>1$, the constant ratio $\Delta B_{\bot}/\Delta B_{\parallel }=1.27$, observed in BNFSO above $\sim \theta$ [inset in Fig.~\ref{fig2}(b)], unambiguously sets the ratio of the anisotropy terms for both models \cite{sup}. The temperature-independent line-width ratio reveals that correlations are isotropic in spin space.

\begin{figure}[b]
\includegraphics[width=1\linewidth, trim=0 220 0 55, clip=true]{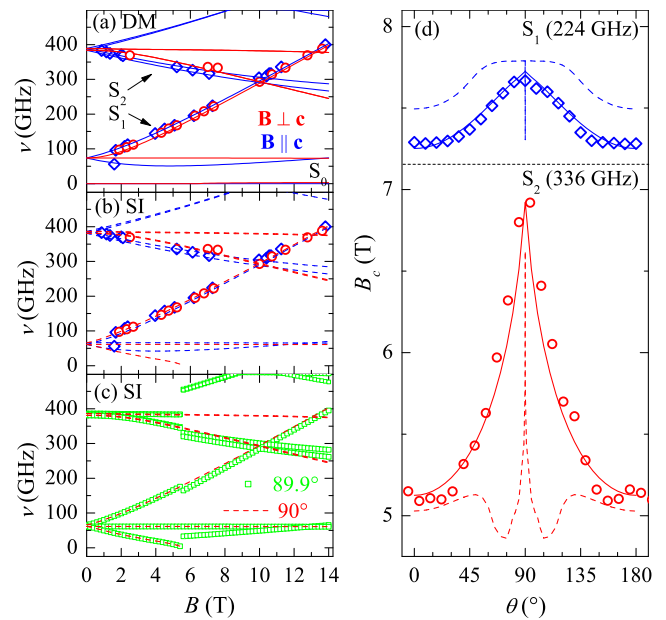}
\caption{(color online). Frequency-field diagram of measured (symbols) and simulated (lines) AFMR modes for (a) DM and (b) SI anisotropy at 4~K. (c) Field-induced phase transition predicted by the SI anisotropy model at 5.5~T for ${\bf B} \bot  {\bf c}$. (d) Angular dependence of the two resonance modes at 4 K; measured (symbols) and simulated for the DM (solid lines) and SI anisotropy (dashed lines).}
\label{fig4}
\end{figure}

In our attempt to determine and quantify the dominant magnetic anisotropy in BNFSO, we now turn to high-frequency measurements. In contrast to X-band, we can follow the ESR signal across $T_N$ [Fig.~\ref{fig3}(a)]. The resonance broadens and shifts to lower field by $\sim 2$~T below $T_N$. The temperature dependence of the line shift mimics an order parameter [Fig.~\ref{fig3}(b)], and corresponds to the opening of a zero-field gap in magnetic excitations (Fig.~\ref{fig4}). This transformation of the paramagnetic signal into a collective antiferromagnetic resonance (AFMR) mode S$_1$ is accompanied by emergence of another AFMR mode (S$_2$) in our frequency window [inset in Fig.~\ref{fig3}(a)]. The frequency-field diagram of the two modes is shown in Fig.~\ref{fig4} for both relevant directions of the applied field with respect to the crystallographic $c$ axis.

We have simulated the AFMR modes experimentally detected at 4~K in a mean-field approximation \cite{sup}, again for both types of magnetic anisotropy [Fig.~\ref{fig4}(a), (b)]. We note that the zero-field gap to the lowest gapped branch S$_1$ is opened by finite anisotropy ($d_c$ or $D$, $a$, $F$) in a square-root fashion \cite{sup}, while the gap to the next branch S$_2$ is present already for isotropic $\mathcal{H}_e$ and only slightly shifts with anisotropy. The latter gap is reproduced with the exchange parameters $J_1=11.9$~K, $J_2=3.4$~K, $J_3=0.72$~K, $J_4=0.24$~K, and $J_5=3.4$~K, which corresponds to a 20\% increase \cite{foot2} of the parameters obtained in the INS study \cite{MartyPRL}. 

The SI model [Eq.~(\ref{eq2})], at first sight, yields a reasonable description of the experimental $\nu-B$ diagram [Fig.~\ref{fig4}(b)] for $D=-17$~mK, $a=-18$~mK and $F=-0.9$~mK \cite{foot3}. However, a closer look at this model unveils its inconsistencies with the experiment. The simulated angular dependence of the two AFMR modes does not match the experiment [Fig.~\ref{fig4}(d)]. The predicted non-monotonic dependence of the S$_2$ mode is a consequence of a field-induced phase transition, occurring around 5.5~T \cite{foot5} for $50^{\circ} <\theta<90^{\circ}$ [see the discontinuity of curves in Fig.~\ref{fig4}(c)]. The apparent absence of the transition at exactly $\theta=90^{\circ}$ is accidental, which can be deduced from Fig.~\ref{fig4}(c). The transition corresponds to the field where the lowest branch reaches zero frequency. Such behavior was indeed observed and justified with the SI model on another Fe-based triangular lattice \cite{Smirnov}. The absence of any irregularities in low-temperature magnetization isotherms speaks against such transition in BNFSO at least up to 23~T \cite{MartyThesis}. We stress that, on the other hand, no such excitation softening exists for the DM model [Fig.~\ref{fig4}(a)]. Finally, the SI model predicts the selection of the $(\epsilon_H,\epsilon_{\Delta})=(-1,1)$ GS independent of the structural chirality. This contradicts with the experimental observation of the $(1,-1)$ GS in a crystal with structural left handedness \cite{MartyPRL}, clearly making the SI model incompetent of selecting the proper ground state. This selection is independent of the structural chirality, because it is a single-ion property. 

The DM model [Eq.~(\ref{eq3})], on the other hand, yields excellent agreement with the experiment for $|d_c|=45$~mK. Both the $\nu -B$ diagram [Fig.~\ref{fig4}(a)] and the finer-detail angular dependence of both AFMR modes [Fig.~\ref{fig4}(d)] are simulated convincingly. The DM anisotropy $d_c$ explains the ESR line width and the finite gap to the S$_1$ branch. Its relative size $|d_c/J_1|=0.4\%$ is in good agreement with the order-of-magnitude prediction $\Delta g/g =0.25\%$, as derived by Moriya \cite{Moriya}. We note that the zero-field gap could also be explained by the presence of symmetric anisotropic exchange (AE) of a similar size $|J^{\rm AE}/J_1|=0.4-0.8\%$ \cite{sup}. However, as the AE anisotropy results from the second order perturbation in $\lambda$ (while DM results from the first order perturbation), it is of the size \cite{Moriya} $(\Delta g/g)^2J_1 =6\cdot 10^{-6}J_1$ and therefore negligible.  We stress that the in-plane DM component $d_{ab}$ does not affect the detected lowest branches and cannot be determined from the studies of these excitations. This makes the above-presented X-band ESR investigation, yielding $d_{ab}/d_c=2.6$, invaluable in determining also the $|d_{ab}|=120$~mK DM component. There is a very good agreement between the excitations observed in ESR and calculated with the DM model on one hand and the modes identified from the INS experiments \cite{Loire,Jensen} on the other hand. The calculated S$_0$ branch [Fig.~\ref{fig4}(a)] is the gapless Goldstone mode emerging at the origin of the reciprocal space (scattering vector $Q$=0) for the correlations of the $c$-axis spin components. The S$_1$ mode corresponds to the other, gapped excitations for chiral correlations of the $ab$-plane spin components at $Q$=0. These excitations at $Q=\pm\bf q$ involve correlations of the $c$-axis spin-component \cite{Loire} and cross $Q$=0 at $\approx 400$~GHz, yielding the S$_2$ signal. 

The DM anisotropy thus proves to be the origin of the unique magnetic properties of BNFSO. Our simulations \cite{sup} further disclose that one of the two possible GS [Fig.~\ref{fig1}(b)] is selected depending on the sign of $d_c$. The experimentally detected GS $(1,-1)$ is realized for $d_c>0$, when the DM energy [Eq.~(\ref{eq3})] is minimized. Formation of a uniform vector chiral state at finite temperature is, however, challenged by formation of domain walls, which prosper on frustrated lattices due to thermal fluctuations \cite{Chalker}. It is likely that these are responsible for diffuse neutron scattering coexisting with LRO in BNFSO far below $T_N$ \cite{ZhouPRB}. It is therefore striking that the minute $d_c=45$~mK term effectively imposes a single-domain GS at two orders of magnitude larger temperatures, as experimentally verified at 1.5~K \cite{MartyPRL}, and a chiral unbalance up to T$_N$, with chiral correlations persisting even in the paramagnetic state \cite{Stock}.

In conclusion, we have shown that the selection of the unique chiral magnetic ground state in BNFSO is due to the Dzyaloshinsky-Moriya interaction as the dominant source of magnetic anisotropy. Although this term is small ($d_c=45$~mK, $d_{ab}=120$~mK), it effectively overcomes thermal fluctuations and leads to a fully chiral state in structurally enantiopure crystals, whereas spin-liquid signatures might also be anticipated above $T_N$, possibly with skyrmion texturation \cite{pappas1,pappas2}.

We thank S. de Brion for fruitful discussion and acknowledge the financial support of the Slovenian Research Agency (projects
J1-2118 and BI-US/09-12-040).

\begin{widetext}
\vspace{19cm}
\begin{center}
{\large {\bf Supplementary information:\\

Role of Antisymmetric Exchange in Selecting Magnetic Chirality in Ba$_3$NbFe$_3$Si$_2$O$_{14}$}}\\
\vspace{0.5cm}
A. Zorko,$^{1,2}$ M. Pregelj,$^1$ A. Poto\v{c}nik,$^1$ J. van Tol,$^3$ A. Ozarowski,$^3$ V. Simonet,$^4$ P. Lejay,$^4$ S. Petit,$^5$ and R. Ballou$^4$
\vspace{0.3cm}

{\it
$^1$Jo\v{z}ef Stefan Institute, Jamova 39, SI-1000 Ljubljana, Slovenia\\
\vspace{0.2cm}
$^2$EN--FIST Centre of Excellence, Dunajska 156, SI-1000 Ljubljana, Slovenia\\
\vspace{0.2cm}
$^3$National High Magnetic Field Laboratory, Florida State University, Tallahassee, Florida 32310\\
\vspace{0.2cm}
$^4$Institut N\'eel, CNRS and Universit\'e Joseph Fourier, BP 166, 38042 Grenoble, France\\
\vspace{0.2cm}
$^4$Laboratoire L\'eon Brillouin, CEA-CNRS, CE-Sacley, F-91191 Gif sur Yvette, France\\
}

\end{center}
\end{widetext}

\section{ESR above $T_N$}

\begin{figure}[b]
\includegraphics[width=0.8\linewidth, trim=0 10 0 10, clip=true]{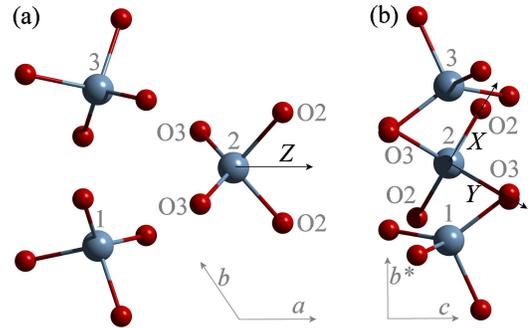}
\caption{(color online). Triangle of magnetically non-equivalent FeO$_4$ tetrahedra in \FeLang~in (a) the $ab$ and (b) the $b^*c$ crystallographic plane. The two crystallographically non-equivalent oxygen sites are labeled as O2 and O3. The principal axes of the $g$-factor on the Fe site 2 are denoted by $X$, $Y$ and $Z$; the latter is parallel to a local two-fold rotational axis.}
\label{fig1s}
\end{figure}

In the linear response theory an electron spin resonance (ESR) spectrum is given by the Fourier transform of the relaxation function \cite{KT}
\begin{equation}
\label{eq1}
\varphi (t)= \frac{\left\langle M^+(t)M^-(0) \right\rangle}{\langle M^+(0)M^-(0)\rangle},
\end{equation}
\noindent where $M^\pm =g\mu_B\sum_{j}S_j^\pm $ are transverse magnetization operators and $\left\langle ... \right\rangle$ denotes canonical averaging. Line position ($g$-value) and line width of the spectrum contain important information on local anisotropies \cite{AB}. The principal axes of the $g$-tensor are set by a local symmetry of the crystal field. In \FeLang~one of the principal axes (labeled $Z$) is parallel to a two-fold rotational axis passing through the center of each FeO$_4$ tetrahedron, while the other two (labeled $Y$ and $X$) are parallel with the two normals to the O$j$-Fe-O$j$ planes, $j=2,3$ labeling two pairs of oxygens at a distance of 1.909~\AA~and 1.856~\AA~from Fe, respectively. For site 2 (see Fig.~\ref{fig1s} and Fig.~1 of the main text) the $Z$ axis corresponds to the crystallographic $a$ direction, while $X$ and $Y$ axes are tilted by $30^{\circ}$ and 120$^{\circ}$ from the $b*$ axis in the crystallographic $b^*c$ plane. The three magnetically non-equivalent Fe sites on each triangle yield a single ESR line due to exchange narrowing. Therefore, an axially symmetric effective $g$-tensor is obtained after averaging over the three sites. We find
\begin{eqnarray}
\label{eq2}%
g_c&=&g_{ZZ}{\rm cos}^2 \alpha +g_{YY}{\rm sin}^2 \alpha,\\
\label{eq3}%
g_{ab}&=&\frac{g_{XX}+g_{YY}{\rm cos}^2 \alpha +g_{ZZ}{\rm sin}^2 \alpha}{2},
\end{eqnarray}
\noindent and $g(\theta)=\sqrt{g_c^2 {\rm cos}^2 \theta + g_{ab}^2 {\rm sin}^2 \theta}$, where  $\theta$ denotes the polar angle between the crystallographic $c$ axis and the applied magnetic field.

When the magnetic anisotropy term $\mathcal{H}'$ of the spin Hamiltonian [see Eq.~(1) in the main text] is inferior to the isotropic exchange term $\mathcal{H}_e$ and/or the Zeeman term $\mathcal{H}_Z$ it can be treated as a perturbation. In cases of moderate line widths ($\Delta B\ll k_B J/g\mu_B$) the well-established Kubo-Tomita approach applies for $T\gtrsim J$. The ESR absorption spectrum then has a Lorentzian line shape and a full-width-half-maximum (FWHM) line width, given at infinite temperatures as
\begin{equation}
\label{eq4}
\Delta B_\infty = \sqrt{2\pi}  \frac{ k_{B}}{g \mu _{B}} \sqrt{\frac{M_{2}^{3}}{M_{4}}}.
\end{equation}
\noindent The second and the fourth moment of the resonance spectrum are given by \cite{Castner}
\begin{eqnarray}
\label{eq5}%
M_{2}&=&\frac{\left\langle \left[ \mathcal{H}^{\prime },S^{+}\right]
[ S^{-},\mathcal{H}^{\prime }] \right\rangle }{\left\langle S^{+}S^{-}\right\rangle
}\; ,
\\%
\label{eq6}%
M_{4}&=&\frac{\left\langle \left[ \mathcal{H}-\mathcal{H}_{Z},\left[
\mathcal{H}^{\prime },S^{+}\right] \right] [ \mathcal{H}-\mathcal{H}_{Z},\left[
\mathcal{H}^{\prime },S^{-}\right] ] \right\rangle }{\left\langle
S^{+}S^{-}\right\rangle }\; ,
\end{eqnarray}
\noindent where $\left[ \; ,\; \right]$ denotes a commutator. The moments given by the spin Hamiltonian are always finite, while those of the Lorentzian line shape are infinite. Therefore, the true experimental line shape deviates from the Lorentzian shape and a modified line shape has to be employed. Typically, this deviation becomes notable only in extreme wings. The criterion for the selection of a suitable line shape should be the agreement between the theoretical exchange field $B_e=k_B/g\mu_B\sqrt{M_4 /M_2}$, determined by
the spin Hamiltonian, and the one given by a particular line shape \cite{Castner}.  We find that in \FeLang~the approximation where the Lorentzian
line shape is multiplied with the broad Gaussian, $G(B)\propto {\rm exp}\big[-\left(B
-B_0\right)^2 /2B_e^2\big]$, yields an excellent agreement of the exchange fields. This situation is usual in systems with strong exchange, where the Gaussian time decay of spin correlations can be assumed. This approximation yields the $\sqrt{2\pi}$ prefactor in Eq~(\ref{eq4}).

Although tedious, calculations of the two moments lead to analytic expressions of the ESR line width. We calculated it separately for the case of the single-ion (SI) anisotropy, taking into account spin operators up to the fourth order  [see Eq.~(2) in the main text], and for the Dzyaloshinsky-Moriya (DM) interaction [see Eq.~(3) in the main text]. For the spin lattice shown in Fig.~1 of the main text we derived the following expressions for the ESR line width

\begin{widetext}
\begin{eqnarray}
\label{eq5}%
\Delta B_\infty^{\rm SI} &=&\sqrt{2\pi}  \frac{ k_{B}}{g \mu _{B}} \sqrt{\frac{27\left[4 D^2 + \frac{32}{7} a^2 +  \frac{10}{9}F^2 + \frac{10}{3} aF  - \frac{2}{45}
\left( 18 D^2 + 5 F^2 + 15 aF\right) {\rm cos}\;2 \theta \right]^3}{20 J^2_{\rm SI} \left[ 189 D^2 + 720a^2 + 175 F^2 + 525 aF - \frac{7}{5} \left( 27D^2 + 25F^2 + 75 aF \right) {\rm cos}\;2 \theta \right]}},\\
\label{eq6}%
\Delta B_\infty^{\rm DM} &=&\sqrt{2\pi}  \frac{ k_{B}}{g \mu _{B}} \sqrt{\frac{105\left[ 5d_{ab}^2+6d_c^2+ \left(d_{ab}^2-2d_c^2 \right) {\rm cos}\;2 \theta \right]^3}{32\left[ 35 J_{\rm DM}^2 d_{ab}^2 +6 J'^2_{\rm DM} d_c^2  + \left( 2 J'^2_{\rm DM} d_c^2  - 7 J_{\rm DM}^2 d_{ab}^2  \right) {\rm cos}\;2 \theta \right]}}.
\end{eqnarray}
\end{widetext}

\noindent Here $J^2_{\rm SI}=J_1^2+2J_2^2+J_3^2+J_4^2+J_5^2$, $J^2_{\rm DM}=3J_1^2+2J_2^2+J_3^2+J_4^2+J_5^2$ and $J'^2_{\rm DM}=18J_1^2+14J_2^2+7J_3^2+7J_4^2+7J_5^2$.

The ratios of the anisotropy parameters within each model are set by the temperature independent experimental line-width ratio $\Delta B_{\bot}/\Delta B_{\parallel }=1.27$. In case of the SI model, our calculations predict $\Delta B_{\bot}/\Delta B_{\parallel }=1$ for the $a$ term and $\Delta B_{\bot}/\Delta B_{\parallel }=1.5$ for the $D$ term. Due to its cubic symmetry the former term does not yield any anisotropy, however in still gives a finite ESR line width, because it does not commute with $\mathcal{H}_Z$. In case of the DM model, the $d_{ab}$ anisotropy leads to $\Delta B_{\bot}/\Delta B_{\parallel }=1.5$ and the $d_c$ anisotropy to $\Delta B_{\bot}/\Delta B_{\parallel }=0.5$. In both models, a single dominant parameter thus fails to account for the experiment.

\section{AFMR below $T_N$}

A mean-field approach is used to calculate the antiferromagnetic resonance (AFMR) modes. In this approximation, sublattices are introduced instead of individually treatment of the spins. The corresponding magnetizations are given by ${\bf M}_j=-Ng\mu_B\langle {\bf S}_j\rangle$, where $N$ is the number of Fe$^{3+}$ magnetic ions in the $j$-th sublattice. The AFMR modes are calculated within the spin Hamiltonian given by Eqs.~(1)-(3) in the main text. Applying the mean-field approximation \cite{Fairall} we can rewrite this Hamiltonian into a magnetic free energy ${\cal F}$ per Fe site,
\begin{widetext}
\begin{eqnarray}
\nonumber
{\cal F}  &=&  \sum_{i>j} \tilde {J}_{ij}{\bf M}_i\cdot{\bf M}_j  +  \mu_B  {\bf B}_0 \cdot {{\bf g}\over g}\cdot\sum_{j}{\bf M}_j + \sum_{i>j}{\bf \tilde {d}}_{ij}\cdot {\bf M}_i\times{\bf M}_j+\\ 
&+& \sum_{j} \left[ \tilde{D}_j M_{j,Z}^2 + \frac{\tilde{a}_j}{6} \left( M_{j,X}^4+M_{j,Y}^4+M_{j,Z}^4\right) +  \frac{\tilde{F}_j}{180} \left( 35M_{j,Z}^4 -M_{j,Z}^2\{30 M^2 -25\}\right)\right] + const.
\end{eqnarray}
\end{widetext}
where the sums run over all sublattices. The  molecular field constants are defined as
\begin{align}
\nonumber \tilde{J}_{ij}  =&  {Z_{ij} J_{ij} \over N (g\mu_B)^2}\, , \, \tilde{\bf d}_{ij} = {Z_{ij} {\bf d}_{ij}\over N (g\mu_B)^2}\, ,\\
\tilde{D}_j  =  {D_j\over N (g\mu_B)^2}\, &,\,
\tilde{F}_j  =  {F_j\over N (g\mu_B)^4}\, ,\,
\tilde{a}_j =  {a_j\over N (g\mu_B)^4}\, ,
\end{align}
where $Z_{ij}$ is the number of the $ij$ neighbors.

The magnetic ground state is obtained from minimization of the free energy. This then yields effective magnetic fields acting on individual sublattice magnetization 
${\bf B}_j=-{\partial {\cal F}\over \partial {\bf M}_j}$, 
which is introduced into the equations of motion for all sublattices
\begin{eqnarray}
{d{\bf M}_j\over dt}=-\frac{g\mu_B}{\hbar}{\bf} M_j\times{\bf B}_j\, .
\end{eqnarray}
To make the problem tractable, harmonic oscillations are assumed, $dM_j/dt\propto \exp\left( i\omega t\right)$, and the equations are linearized. 
The problem is solved numerically -- the parameters $\tilde{J}_{ij}$, $\tilde{{\bf d}}_{ij}$, $\tilde{D}_j$, $\tilde{a}_j$ and $\tilde{F}_j$ are adjusted to fit the field- and the angular dependence of the antiferromagnetic resonance modes.

In this mean-field approach all information about the interactions between spins within a particular sublattice is lost. Therefore, a meaningful choice of sublattices is of great importance. The ferro-triangular magnetic order allows a restriction to six sublattices in each triangular plane, in order to account for the lowest lying magnon modes (see Fig.~1 in the main text for the site numbering). Thus both the intratriangle exchange $J_1$ and the intertriangle exchange $J_2$ are taken into account. The effect of the latter is however simplified, since all the 1'-3' triangles neighboring the basic 1-3 triangle are assumed the same, i.e., effectively, only one 1-3 and one 1'-3' triangle with the inter-triangle exchange $J_2$ are needed. Finally, to take into the account also the interlayer interactions ($J_3$, $J_4$ and $J_5$), we stacked 14 pairs of 1-3 and 1'-3' triangles and used periodic boundary conditions, thus allowing for the experimental pitch of 2$\pi$/7 along the $c$-axis.
\begin{figure}[t]
\includegraphics[width=0.95\linewidth, trim=40 10 50 35, clip=true]{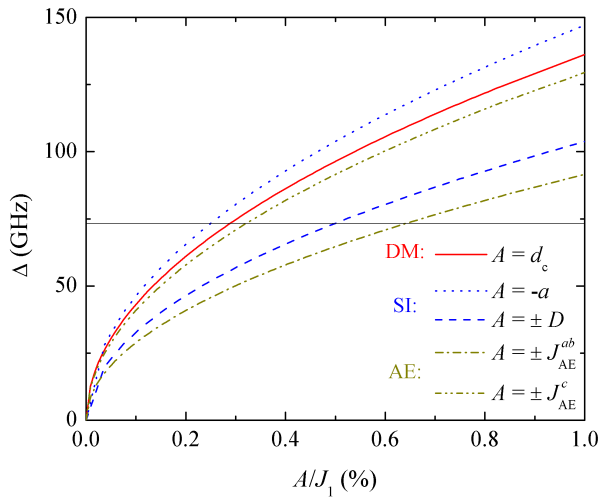}
\caption{(color online). Dependence of the energy gap to the lowest gapped branch on the anisotropy size $A$ for the Dzyaloshinsky-Moriya (DM), single-ion (SI) anisotropy and symmetric anisotropic exchange (AE). The horizontal line corresponds to the experimentally observed gap.}
\label{fig2s}
\end{figure}

The resulting magnetic ground state is a helix with the ordering vector ${\bf q}=(0,0,1/7)$. The plane of the helix will, however, be randomly oriented if $\mathcal{H}'$ is set to zero, thus resulting in a sinusoidal tilt (along $c$) of spins on a given site from the $c$ axis. This remains the case if a finite in-plane DM component $d_{ab}$ is present. Adding a finite out-of-plane DM component $d_c$ aligns the helix plane with the crystallographic $ab$ plane. The triangular chirality is found anti-parallel to the ${\bf d}_c$ vector, thus minimizing the DM interaction energy. The helicity is always anti-parallel to the triangular chirality due to the left-handedness of the crystal structure imposing $J_5>J_3$. In the SI model, on the other hand, $a>0$ leaves the helix plane randomly oriented, while $D>0$ randomly imposes helix planes with their normals perpendicular to the $c$ axis. On the other hand, spins are bound to $ab$ planes for $D<0$ and/or $a<0$, which, however, always select the state with positive triangular chirality (and negative helicity).  

The AFMR modes that suit the experimental observations best are plotted in Fig.~4 of the main text for both anisotropy models. In zero field, the three lowest modes define the lowest two branches S$_0$ and S$_1$, which are characterized by $\sum_{\Delta}{\bf S}_j=0$ and a linear motion of spins. The crucial difference between the two anisotropy models is that in the DM model the S$_0$ mode is the gapless Goldstone mode and the two S$_1$ modes occur at ($\Delta=73$~GHz). In the SI model, on the other hand, all three modes are gapped -- a two-fold degenerate mode is found at $\Delta=61$~GHz and the third mode at $\Delta=66$~GHz. The small splitting is due to the fact that the SI anisotropy in \FeLang~is multi-axial (see Fig.~1 of the main text). If the magnetic helix was incommensurate a gapless Goldstone mode would exist also in the SI model, as there would be no energy cost in rotating the helix around its axis. The gapless mode is, however, absent for a commensurate helix. Experiments imply the latter. Namely, no changes in the propagation vector, which is a rational fraction of $c$ with small numerator and denominator co-primes, were detected in the temperature range of the ordered phase \cite{MartyThesis}. This is not in line with the incommensurability, arising from a delicate balance between the exchange parameters $J_3-J_5$, then subject to strong sensitivity on temperature \cite{Pregelj}. This then justifies the use of the finite lattice of 14 sites along the $c$ axis in our modeling.

The zero-field gap to the lowest gapped mode opens in a square-root fashion (Fig.~\ref{fig2s}) for both the DM and the SI model. We have calculated the dependence of the gap also for symmetric anisotropic exchange (AE)
\begin{equation}
\mathcal{H'}_{\rm AE} = \sum _{(ij)}{J_{\rm AE}^{\xi }S_i^{\xi}S_j^{\xi}},
\end{equation}
\noindent where the sum runs over the nearest-neighbor Fe pairs, and $\xi$ either corresponds to crystallographic $c$ direction or a general local direction with respect to the $ij$ bond within the $ab$ plane. Similarly to other two models, in the AE model the square-root dependence of the gap is observed. Moreover, all models predict a very similar size of the anisotropy $A$, which is required to explain the gap.

\appendix


\begin{thebibliography}{000}
\bibitem{foot1} Dipolar interaction between Fe$^{3+}$ moments and their hyperfine coupling to nuclear moments are negligible -- they yield line widths of $\Delta B \lesssim 3$~mT due to strong exchange narrowing. Symmetric anisotropic exchange being a quadratic correction in $\lambda$ is inferior to the antisymmetric Dzyaloshinsky-Moriya term being linear in $\lambda$ \cite{Moriya}.
\bibitem{foot2} The increase of the exchange parameters by $20\%$ compared to those from the INS experiment \cite{MartyPRL} is necessary because we use the experimental value of the ordered moments 4$\mu_B$ instead of the full value 5$\mu_B$.
\bibitem{foot3} Other selections of the sign for the SI parameters do not yield AFMR modes resembling the experiment.
\bibitem{foot5} The critical field is shifted to 7.5~T if exchange parameters from Ref.~\cite{Stock}, which are significantly different, are used.
\end{thebibliography}
\end{document}